\documentclass[aps,prl,final,twocolumn,letterpaper]{revtex4-1}

\usepackage{graphicx}   
\usepackage{import}                         
\usepackage{epstopdf}
\usepackage{amsmath} 
\usepackage{float} 
\usepackage{bm}
\usepackage{amssymb}
\usepackage{quotes}
\usepackage{indentfirst}
\usepackage{color}
\usepackage{transparent}
\usepackage{dcolumn}
\usepackage{braket}
\usepackage{multirow}
\usepackage{cancel} 
\usepackage{mdframed}
\usepackage{color}
\usepackage{bm}
\usepackage{dsfont}
\usepackage{slashed}
\usepackage{soul, color} 
\soulregister\ref{7}  
\soulregister\cite{7} 
\renewcommand{\st}[1]{}

\usepackage{xr}

\makeatletter
\newcommand*{\addFileDependency}[1]{
  \typeout{(#1)}
  \@addtofilelist{#1}
  \IfFileExists{#1}{}{\typeout{No file #1.}}
}
\makeatother

\newcommand*{\myexternaldocument}[1]{%
    \externaldocument{#1}%
    \addFileDependency{#1.tex}%
    \addFileDependency{#1.aux}%
}

\myexternaldocument{si}

\usepackage{textcomp} 
\usepackage{xifthen}
\usepackage{xcolor}
\usepackage{etoolbox}
\newboolean{togglechanges} 
\usepackage{siunitx}
\usepackage{wrapfig}

\setboolean{togglechanges}{false}

\newcommand{\comment}[1]{\ifbool{togglechanges}
    {#1}  
    {\textcolor{blue}{#1}}}

\usepackage{bibentry}

\usepackage{graphicx}
\usepackage{dcolumn}
\usepackage{bm}
\setcitestyle{numbers,square,super}

\usepackage{textcomp} 
\begin{document}
\rmfamily

\title{Vacuum Fluctuation-Induced State Switching in Degenerate Optical Parametric Oscillators}
\author{Yihao~Huang$^{1,\ddag}$}
\author{Seou~Choi$^{2,\ddag}$}
\email{seouc130@mit.edu}
\author{Rom Simovitch$^{3}$}
\author{Jamison Sloan$^{2,4}$}
\author{Charles~Roques-Carmes$^{2,4}$}
\author{Michael Horodynski$^{1,2}$}
\author{Marin~Solja\v{c}i\'{c}$^{1,2}$}
\author{Yannick Salamin$^{1,2,3}$}
\email{yannick.salamin@ucf.edu}
\affiliation{$^\ddag$ denotes equal contribution.\looseness=-1}
\affiliation{$^{1}$ Department of Physics, Massachusetts Institute of Technology, Cambridge, MA, USA\looseness=-1}
\affiliation{$^{2}$ Research Laboratory of Electronics, Massachusetts Institute of Technology, Cambridge, MA, USA\looseness=-1}
\affiliation{$^3$ CREOL, The College of Optics and Photonics, University of Central Florida, Orlando, Florida, USA\looseness=-1}
\affiliation{$^{4}$ E. L. Ginzton Laboratories, Stanford University, 348 Via Pueblo, Stanford, CA, USA\looseness=-1}

\clearpage 

\setlength{\parskip}{0em}
\vspace*{-2em}


\vspace{0.8cm}

\begin{abstract}

Bistable driven–dissipative systems near bifurcations can exhibit noise-activated switching between steady states. Here, we investigate how quantum vacuum fluctuations induce  such switching in a biased optical parametric oscillator (OPO), a nonlinear system with intrinsic bistability. We show how microscopic quantum fluctuations driving macroscopic transitions can be controlled with an external bias field that reshapes the OPO steady-state metapotential. We derive  analytical expressions for the average switching time and validate them through simulations of the OPO field distribution and inter-state probability flow under bias injection. We further examine how switching depends on bias strength, pump gain, and optical nonlinearity. Our findings clarify how quantum noise can shape macroscopic dynamics and provide a foundation for noise-assisted photonic machine learning and probabilistic quantum gates.

\end{abstract}

\maketitle

\section*{Introduction} 

The coexistence of multiple stable states in driven, interacting systems is ubiquitous across chemistry, physics, and biology. Noise-activated switching between coexisting macroscopic states governs chemical reaction rates~\cite{kramers1940brownian,hanggi1990reaction}, switching in condensed matter~\cite{lapidus1999stochastic, chan2007activation,camsari2017implementing}, and perceptual bistability in neural circuits~\cite{roxin2008neurobiological,shpiro2009balance}. In such systems, the fluctuation-induced switching behavior is governed by the height of an effective activation barrier in a potential landscape, whose minima represent metastable states. External driving can reshape this landscape, enabling control over the activation barrier height and switching rates~\cite{siddiqi2004rf,borders2019integer,kaiser2022hardware}.

In driven quantum systems, switching dynamics depart from classical thermally activated diffusion and are instead governed by intrinsic quantum fluctuations\cite{dykman1998fluctuational, marthaler2006switching, boness2024resonant, su_unraveling_2025}. Superconducting microwave resonators provide a prominent example, where quantum activation, tunneling, and coherent effects can govern transitions between metastable states \cite{andersen2020quantum, de2026asymmetry}. These studies establish switching as a sensitive probe of fluctuations across both classical and quantum regimes, while highlighting the central role of the underlying noise source and system nonlinearity.

Although controllable switching driven by quantum vacuum fluctuations would provide a powerful route to manipulating dynamical phases of quantum optical systems, it remains elusive for most driven-dissipative platforms. Systems with macroscopic steady states, such as above-threshold optical parametric oscillators (OPOs) and lasers, possess potential landscapes with large effective activation barriers, leading to exponentially long switching times that exceed experimentally accessible observation windows. Operating near threshold lowers these barriers but replaces controlled, directional transitions with symmetric, undirected fluctuations between metastable states.

Here, we show that an external coherent bias can reshape the non-equilibrium potential landscape of driven-dissipative systems operating in the above-threshold regime, reducing the activation barrier and enabling observable  stochastic switching between metastable states driven by quantum vacuum fluctuations. We demonstrate our concept in a degenerate OPO, a representative driven-dissipative system exhibiting bistability between phase-degenerate coherent states arising from spontaneous symmetry breaking. By mapping the system dynamics to an effective metapotential, we derive analytical expressions for the switching time and show that it depends exponentially on the bias-controlled barrier height. These predictions are validated using stochastic simulations of the underlying Fokker–Planck dynamics. We systematically examine how the switching behavior depends on the bias field strength, pump gain level, and optical nonlinearity. Our results establish a framework for engineering fluctuation-driven switching in quantum optical systems, providing direct access to quantum activation dynamics in a regime where vacuum fluctuations dominate, and opening a route toward tunable probabilistic behavior in photonic platforms.   

\begin{figure}
    \centering
    \includegraphics[scale = 1.0]{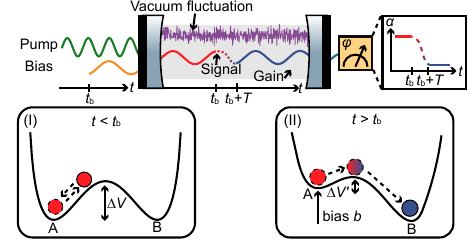}
    \vspace*{-4mm}
    \caption{\textbf{Optical switching seeded by quantum vacuum fluctuations.} Driven-dissipative quantum optical systems such as the optical parametric oscillator (OPO) can exhibit quantum vacuum-fluctuation-induced switching between metastable states. Without the change in the potential barrier (i.e., in the absence of the bias field that reshapes the potential landscape), the large effective potential barrier makes switching unobservable on realistic timescale. The bias field can lift up one of the potential wells and control the switching rate between bistable OPO states induced by quantum vacuum fluctuations.}
    \label{fig:concept}
\end{figure}

\section*{Results} 

Fig.~\ref{fig:concept} demonstrates the mechanism of quantum vacuum fluctuation-induced switching within an OPO. The system consists of a parametric gain medium embedded in an optical cavity. In this study, we consider a degenerate OPO (DOPO), in which a pump field at frequency $2\omega$ undergoes parametric down-conversion to generate signal and idler fields at the same frequency $\omega$. 

The temporal evolution of the cavity signal mode under an injected bias field $b$ is governed by the Heisenberg-Langevin equation~\cite{roques2023biasing}:
\begin{equation}
\begin{aligned}\label{eq:Langevin}
    \dot{a} = -a + \lambda a^\dagger - g^2(a^\dagger a)a + b+ F(t).
\end{aligned}
\end{equation}
where $a$ is the annihilation operator for the DOPO signal. The dynamics are determined by the fraction of the pump field above the threshold $\lambda$, the quantum noise level $g$ ($g^2$ is the optical nonlinearity divided by the cavity decay rate), and the Langevin noise operator $F(t)$. 

To analyze the stochastic dynamics, we map the Heisenberg-Langevin equation onto a Fokker-Planck equation using the positive $P$ representation. This formulation describes the quasi-probability distribution of the DOPO signal $P(\alpha,\beta)$ within a complex phase space ($\alpha,\beta$), where $\alpha$ and $\beta$ are independent complex variables corresponding to the signal-mode quadrature. The resulting distribution represents a statistical ensemble of the signal's stochastic trajectories~\cite{carmichael2007statistical}.

\begin{equation} \begin{aligned}\label{eq:Fokker} \frac{\partial P(\alpha,\beta)}{\partial \tau} = \bigg( &-\partial_\alpha A_{\alpha,\beta} -\partial_\beta A_{\beta,\alpha} \\ &+\frac{1}{2}\partial_\alpha^2 B_\alpha^2 +\frac{1}{2}\partial_\beta^2 B_\beta^2 \bigg)P(\alpha,\beta) . \end{aligned} \end{equation}
where $A_{X,Y}=-X+b+Y(\lambda-g^2X^2)$ and $B_{X} = \sqrt{\lambda-g^2X^2}$ denote the drift and diffusion terms, respectively. $\tau$ is a dimensionless time defined as $\tau \equiv \gamma t$, where $\gamma$ is the cavity decay rate. A detailed derivation of the Fokker-Planck equation can be found in Supplementary Note S1.

To characterize the non-equilibrium landscape of the stochastic system, we introduce a metapotential $V$. For systems subject to a constant diffusion term, the drift term $A$ and the metapotential $V$ satisfy the gradient relationship: $A_{X,Y} = -\nabla V(X,Y)$. To transform the DOPO dynamics into a coordinate system with constant diffusion, we perform the coordinate transformation $(\alpha,\beta) \rightarrow (u,v)$ defined by: $u=\textrm{arcsin}\frac{g\alpha}{\sqrt{\lambda}}+\textrm{arcsin}\frac{g\beta}{\sqrt{\lambda}} $ and $v=\textrm{arcsin}\frac{g\alpha}{\sqrt{\lambda}}-\textrm{arcsin}\frac{g\beta}{\sqrt{\lambda}}$. In this transformed coordinate, the Fokker-Planck equation can be rewritten in terms of the metapotential $V$, providing direct access to the system's steady-state distribution. The closed analytical form of the metapotential $V$ and the Fokker-Planck equation after the coordinate transformation can be rewritten as below:

\begin{equation}
\begin{aligned}\label{eq:Fokker_no_noise}
    V(u,v) &= \lambda (\cos u - \cos v)+ \left(\frac{2gb}{\sqrt{\lambda}}-2\right)\ln(\cos u + \cos v) \\
    &-\frac{2gb}{\sqrt{\lambda}}\ln\Bigg(2+2\sin\left(\frac{u+v}{2}\right)+2\sin\left(\frac{u-v}{2}\right)\\
    &+ \cos v-\cos u\Bigg)_{\textstyle ,}\\
    \frac{\partial P(u,v)}{\partial \tau} &= -P\nabla V   - g^2\nabla P.
\end{aligned}
\end{equation}

The steady-state probability distribution in the transformed coordinate $P(u,v)$ follows a Boltzmann-like distribution $P_{\textrm{ss}}(u,v) \propto e^{-V/g^2}$, where the constant diffusion term $g^2$ acts as an effective temperature that spreads the probability across the landscape. To describe the non-stationary probability distribution  $P(u,v)$ during the switching, an additional pre-factor $\beta(u,v)$ is included, yielding $P(u,v)=\beta(u,v)e^{-V/g^2}$. While the pre-factor $\beta(u,v)$ at $t=0$ is determined by the initial probability distribution,  the switching dynamics is encoded in the time-evolution of the pre-factor $\beta(u,v)$. A detailed derivation of the metapotential $V(u,v)$ can be found in the Supplementary Note S2.

Panel I of Fig.~\ref{fig:concept} illustrates the cross-section of the metapotential along the $u$-axis $V(u,v=0)$ for a DOPO pumped above threshold. The system exhibits a symmetric bistable potential, reflecting the energy degeneracy of two coherent steady states with opposite phases (0 and $\pi$ phase). These two coherent states are separated by a potential barrier $\Delta V$. While quantum vacuum fluctuation induces local perturbations near the steady states, the weak intrinsic optical nonlinearity ($g^2 \ll \lambda-1$) results in a high potential barrier that suppresses switching. Therefore, the switching time in the DOPO significantly exceeds the experimentally accessible observation windows.

We now describe how the switching dynamics of the DOPO can be controlled by introducing a strong coherent bias field at time $t=t_{\textrm{b}}$ (panel II of Fig.~\ref{fig:concept}). The bias field introduces an additional drift term that reshapes the metapotential landscape. By adding a bias field that compensates the drift term near one of the steady states, the corresponding potential well is elevated, which reduces the height of the effective potential barrier $\Delta V'$. In this scenario, quantum vacuum fluctuations provide sufficient momentum to overcome the reduced barrier and induce switching between steady states. 

We then derive the switching time of the DOPO. The switching time $T_{\textrm{A} \rightarrow \textrm{B}}$ is defined as a mean time required for the probability distribution of the OPO steady state to move from the initial steady state (A) to the other steady state (B). 
\begin{equation}
\begin{aligned}\label{eq:switching_time_Def}
    T_{\textrm{A} \rightarrow \textrm{B}} = \frac{P_\textrm{A}}{I_{\textrm{A} \rightarrow \textrm{B}}},
\end{aligned}
\end{equation}
where $P_{\textrm{A}}$ is the probability mass at the steady state A and $I_{\textrm{A} \rightarrow \textrm{B}}$ is the probability flux from A to B. $P_{\textrm{A}}$ is obtained by integrating the probability distribution $P(u,v)$ over the region enclosing the potential well of steady state A. $I_{\textrm{A} \rightarrow \textrm{B}}$ can be evaluated by integrating the current density $\textbf{J}$ along a cross-section separating the two steady states. From the continuity equation $\frac{\partial P}{\partial t} +\nabla \cdot \textbf{J} = 0 $ and Eq.~\ref{eq:Fokker_no_noise}, the current density takes the form $\textbf{J}=-g^2\nabla\beta e^{-V/g^2}$. To derive a closed-form analytical solution for $T_{\textrm{A} \rightarrow \textrm{B}}$, we evaluate $P_{\textrm{A}}$ and $I_{\textrm{A} \rightarrow \textrm{B}}$ under two approximations. First, the switching time is dominated by the duration required to reach the saddle point. Since the drift term near the steady state B substantially exceeds that near the steady state A (i.e., the potential well with the steady state B is significantly deeper (panel II of Fig.~\ref{fig:concept})), the probability mass can ``sink'' into the steady state B instantaneously after overcoming the barrier. Therefore, the transit time from the saddle point to the steady state B is neglected in Eq.~\ref{eq:switching_time_Def}. Second, the metapotential landscape near the steady state A is approximated using a finite polynomial expansion. In this work, the metapotential is approximated via a second-order polynomial expansion, while a detailed analysis regarding the order of polynomial approximation and its convergence with simulated data is discussed in the Supplementary Note S3. Under these approximations, the switching time $T_{\textrm{A} \rightarrow \textrm{B}}$ is given by

\begin{equation}
\begin{aligned}\label{eq:Switching time}
    T_{\textrm{A} \rightarrow \textrm{B}} \sim \frac{e^{\Delta V'/g^2}}{\gamma}\frac{\pi}{\lambda -1}\sqrt{1+\frac{1}{\lambda}},
\end{aligned}
\end{equation}

where $\Delta V'$ is the potential barrier height shown in the panel II of Fig.~\ref{fig:concept}, defined as a function of the OPO system parameters and the bias field. $\Delta V'$ is given by the difference between the metapotential at steady state A and the saddle point. As the bias field $b$ increases, $\Delta V'$ decreases until the bias is strong enough to confine the probability distribution to one of the steady states (i.e., the metapotential becomes monostable). A detailed derivation of the switching time can be found in the Supplementary Note S3.

\begin{figure}
    \centering
    \includegraphics[scale =1.0]{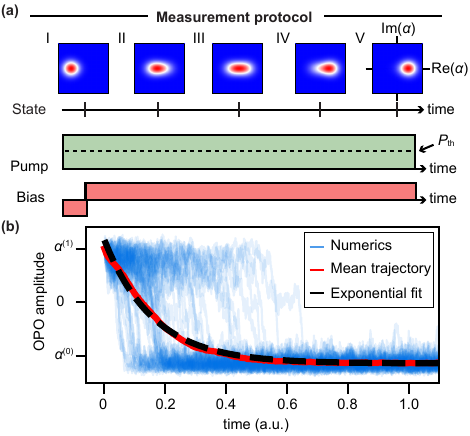}
    \vspace*{-4mm}
    \caption{\textbf{Switching rate estimation protocol of the biased OPO system.} \textbf{(a)} Measurement protocol for the switching rate. The system is initialized such that the OPO probability distribution exists at one of the steady states (panel I), which then switches to the other steady state by injecting a strong positive bias field. \textbf{(b)} Stochastic trajectories of the OPO signal showing the switching behavior under a strong bias field. To determine the switching time, 100 independent trajectories are numerically simulated to obtain the mean trajectory. $\lambda=2.0$, $g=0.4$, and $b=0.9$ in (b).}
    \label{fig:theory}
\end{figure}

The protocol for extracting the switching time of the biased DOPO is illustrated in Fig.~\ref{fig:theory}. For simplicity, the DOPO is initialized in one of the steady states (e.g., $P_{\textrm{A}}=1$) as shown in panel I of Fig.~\ref{fig:theory}(a). The initialization is achieved by pumping the DOPO above threshold while injecting a strong bias to make the system monostable. The sign of the bias is then reversed, reshaping the metapotential so that the initially occupied state becomes metastable with a shallow potential barrier. Under these conditions, the probability distribution will diffuse toward the B side until the entire probability mass is trapped within the B side of the metapotential as shown in panel V of Fig.~\ref{fig:theory}(a) (i.e., $P_{\textrm{A}}=0$). The switching time is extracted as the characteristic time required for the probability mass to transfer from the initial state to the opposite state.

\begin{figure}
    \centering
    \includegraphics[scale = 1.0]{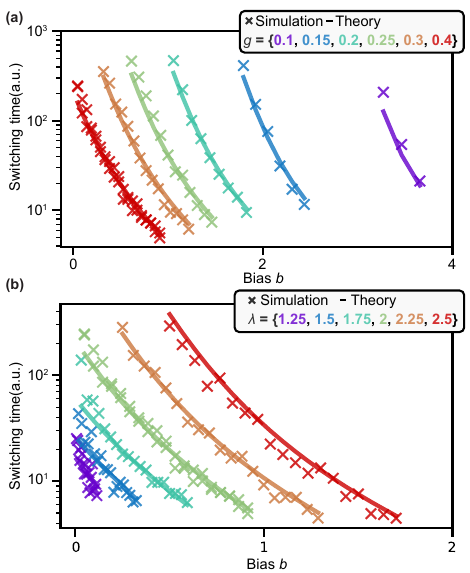}
    \vspace*{-4mm}
    \caption{\textbf{Dependence of switching times on system parameters.} Dependence of switching times on the \textbf{(a)} quantum noise level $g$ with a constant pump strength ratio $\lambda = 2.0$ and \textbf{(b)} pump strength ratio $\lambda$ with a constant quantum noise level $g=0.4$. 100 independent stochastic trajectories are sampled to calculate the switching time at each system parameter.}
    \label{fig:switching_times}
\end{figure}

Fig.~\ref{fig:theory}(b) illustrates the switching dynamics of the DOPO under a strong bias field. Individual trajectories are sampled by numerically solving the Fokker-Planck equation of the DOPO starting from the steady state A, which eventually reaches steady state B. Although the stochastic nature of quantum vacuum fluctuations causes the switching time of individual trajectories (blue curves) to vary, the mean trajectory (red curve) exhibits an exponentially decaying behavior. The switching time is obtained as the decay constant of an exponential fit (black curve) to the mean trajectory.

Leveraging the protocol described in Fig.~\ref{fig:theory}, we examine the dependence of the switching time on various system parameters (Fig.~\ref{fig:switching_times}). Figure~\ref{fig:switching_times}(a) shows that the switching time increases as $g$ decreases. Since $g$ characterizes the effective optical nonlinearity and quantum noise scale, a larger $g$ strengthens the nonlinear saturation term, reducing the effective drift $A_{X,Y}$. This manifests as a shallower metapotential well and therefore a shorter switching time. Figure~\ref{fig:switching_times}(b) shows that the switching time increases with the pump field strength. A stronger pump field leads to a faster amplification of the OPO signal and deepens the potential well, causing the probability mass to remain more strongly trapped in one of the steady states. In both cases, increasing the bias field reduces the switching time by lowering the effective activation barrier. The resulting exponential dependence of the switching time on bias in Fig.~\ref{fig:switching_times} demonstrates that optical bistability in the DOPO can be precisely controlled by an external bias field. 

\begin{figure}
    \centering
    \includegraphics[scale = 1.0]{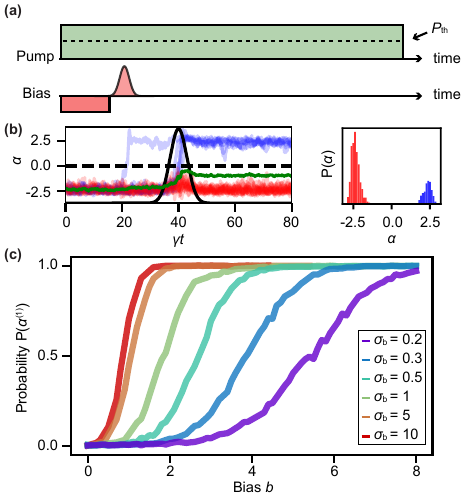}
    \vspace*{-4mm}
    \caption{\textbf{Switching dynamics of the OPO under a pulsed bias field.} \textbf{(a)} Measurement protocol under a pulsed bias field. \textbf{(b)} Left panel: Trajectories of OPO signals (blue and red curves) under a pulsed bias field (black curve). Blue (red) curves denote trajectories that exhibit switching (no switching) respectively, and the average trajectory is shown in green. Right panel: Probability distribution of the OPO signal trajectories at the steady state. \textbf{(c)} Probability of the original OPO signal changing its steady state under different peak bias field levels $b$ and standard deviations $\sigma_{\textrm{b}}$. $\lambda=1.8$ and $g=0.4$ for all the simulations in this figure.}
    \label{fig:pulsed_bias}
\end{figure}

We next consider whether switching can be triggered transiently using a Gaussian pulsed bias field, rather than maintained by a constant bias. Figure~\ref{fig:pulsed_bias}(a) shows the temporal sequence of the pump and bias field injected into the OPO. The OPO probability mass is first initialized  in one phase state by applying a strong bias field, which creates an effectively monostable potential well and localizes the probability distribution.  Then a bias pulse of opposite sign is introduced, transiently reshaping the metapotential and allowing a fraction of the stochastic trajectories to switch to the opposite steady states. Figure~\ref{fig:pulsed_bias}(b) shows the stochastic trajectories of the OPO signal under a pulsed bias field excitation (left panel) and the probability distribution of the trajectories at the steady states (right panel). While the majority of the OPO signals remain in the original well, a subset of trajectories switches to the opposite steady states. This occurs because stochastic trajectories that are close to the saddle points (i.e., $\alpha \sim 0$) gain additional momentum toward the opposite steady state during the bias pulse. After the pulse, the system reverts to an unbiased DOPO configuration, where the activation barrier is large and no further switching is observed.

Figure~\ref{fig:pulsed_bias}(c) illustrates the switching probability $P(\alpha^{(1)})$ as a function of the bias pulse's peak amplitude $b$ and standard deviation $\sigma_{\text{b}}$. The standard deviation $\sigma_{\text{b}}$ is normalized to the cavity lifetime. As expected, the switching probability increases monotonically with both the amplitude and the duration of the bias pulse. Notably, for a fixed peak amplitude $b$, the switching probability $P(\alpha^{(1)})$ changes rapidly as $\sigma_{\text{b}}$ varies within a single cavity lifetime. 

\section{Discussion}

The presented work shows how quantum vacuum fluctuations can induce switching between the bistable steady states of an OPO when its metapotential landscape is reshaped by an external bias. Exploiting the inherent sensitivity of parametric amplification allows microscopic quantum-fluctuation-induced diffusion to drive the stochastic evolution of trajectories across the barrier, converting microscopic quantum fluctuations into observable transitions between macroscopic optical steady states.

The stochastic nature of the switching behavior suggests a pathway for implementing asynchronous probabilistic computing hardware on photonic platforms. Probabilistic computing leverages a probabilistic bit ($p$-bit) that generates 0 or 1 stochastically according to a tunable probability distribution~\cite{kaiser2021probabilistic,camsari2019p,inagaki2021collective}. While various multistable systems leverage thermal fluctuation as a noise source for solving complex optimization problems~\cite{borders2019integer}, recent studies have shown that quantum vacuum fluctuations in optical systems can serve as an intrinsic random source for photonic probabilistic computing~\cite{choi2024photonic,horodynski2025stochastic}. By modulating the bias field in real time, the switching probability can be controlled, implementing various probabilistic machine learning algorithms on photonic hardware. 

The experimental demonstration of the switching behavior requires two capabilities for controlling the bias field: precise control of its amplitude and fast modulation. The strong exponential dependence of the switching time on the bias field indicates that the system is highly sensitive to bias field fluctuations. Increasing $\lambda$ makes the potential valley of the DOPO deeper, which results in the OPO reacting less sensitively to the change in the bias field, making experiments more feasible (Fig.~\ref{fig:switching_times}(b)). Given that integrated platforms support lower pump thresholds owing to smaller mode volumes\cite{lu2025photonic}, the integrated OPO platforms would be better for short-term experimental demonstration compared to the free-space OPO platforms. 

Fast bias modulation is also important. In the idealized protocol considered here, the system is initialized in one phase state using a strong negative bias, after which the bias is rapidly changed to either a constant positive value or a pulsed waveform. If this transition occurs on a timescale comparable to or longer than the cavity lifetime, the metapotential evolves gradually from a monostable to an asymmetric bistable landscape. During this transient evolution, both wells can become shallow, allowing premature switching before the target bias condition is reached. Such non-instantaneous bias modulation would lead to an apparent switching time that differs from the idealized value predicted by the sudden-switching model. 

Although this work focuses on a degenerate OPO, the framework can be generalized to other driven-dissipative systems exhibiting multistable steady states with spontaneous symmetry breaking. For example, a Kerr-nonlinearity-based OPO can also exhibit above-threshold bistability~\cite{gu2025quantum}. In such systems, however, the Kerr-nonlinearity induces a rotation in phase space, making the stochastic dynamics intrinsically two-dimensional, requiring analysis of probability flow over the full metapotential landscape rather than along the one-dimensional cross-section used here. Extending the present framework to these systems could enable a broader class of quantum-noise-driven switching devices in integrated nonlinear photonics.

In conclusion, we have developed a theory of bias-controlled stochastic switching between macroscopic optical steady states mediated by quantum vacuum fluctuations. By mapping the DOPO dynamics to an effective metapotential, we show that the switching behavior follows the diffusion model over a bias-controlled activation barrier. The switching time was obtained both analytically and through stochastic numerical simulations, revealing its dependence on key OPO parameters, including the bias field, pump strength, and nonlinear coupling. The resulting dynamics provide a direct link between microscopic quantum noise and macroscopic optical state selection. This work establishes biased OPOs as a platform for studying quantum activation in optical systems and points toward photonic devices whose probabilistic behavior is controlled by coherent bias fields rather than by externally imposed classical noise.

\section{Data and code availability statement}

The data and codes that are used throughout the study are available from the corresponding authors upon request. Correspondence and requests should be addressed to S.~C. (seouc130@mit.edu) and Y.~S. (yannick.salamin@ucf.edu).

\section{Authors contributions}

S.~C., Y.~S., J.~S., C.~R.-C., and M.~S. conceived the original idea. Y.~H. developed the theoretical and numerical tools, with contributions from S.~C., R.~S., Y.~S., J.~S. The manuscript was written by Y.~H., S.~C., Y.~S., R.~S., and with inputs from all authors.    

\section{Competing interests}

The authors declare no competing interests.

\section{Acknowledgements}

S.~C. acknowledges support from Korea Foundation for Advanced Studies Overseas PhD Scholarship. The authors acknowledge the MIT SuperCloud and Lincoln Laboratory Supercomputing Center for providing computation resources that were essential to the results presented in this paper. This research was funded in part (M.H.) by the Austrian Science Fund (FWF) [10.55776/J4729].  This material is based upon work also supported in part by the U.S. Army Research Office through the Institute for Soldier Nanotechnologies at MIT, under Collaborative Agreement Number W911NF-23-2-0121. This research was supported in part by a Ralph E. Powe Junior Faculty Enhancement Award from Oak Ridge Associated Universities.

\bibliographystyle{unsrt}

\bibliography{bibliography.bib}

\end{document}